\documentclass[prd,preprint]{revtex4}

\usepackage{graphicx}
\usepackage{latexsym}
\usepackage[center]{subfigure}

\begin{document}

\title{Musings on Firewalls and the Information Paradox}
\author{Michael Devin}

\begin{abstract}
	The past year has seen an explosion of new and old ideas about black hole physics.
 Prior to the firewall paper, the dominant picture was the thermofield model apparently implied by ADS/CFT duality\cite{mal2}.
While some seek a narrow response to Almheiri, Marolf, Polchinski, and Sully,(AMPS)\cite{amps}, there are a number of competing models. One problem in the field is the ambiguity of the competing proposals. Some are equivalent while others incompatible. This paper will attempt to define and classify a few models representative of the current discussions. 
\end{abstract}

\maketitle 

\section{Information loss models}

	Hawking set out early on to outline an information loss model\cite{hawk5}, which drew some significant criticism\cite{sus1}. The full consequences of information loss scenarios are still unknown. It is easy for non-unitary theories to lose causality, though energy conservation can still be maintained with care\cite{hu}.  This category is characterized by a many-to-one map of states, as well as an injective map from pure to mixed states. The no-hair theorem is embodied in the first property, in which a nonunitary quantum bleaching operator is applied. The second property is shared by some other theories of fundamental decoherence\cite{hu}. \\

\begin{figure}
\includegraphics[scale=.3]{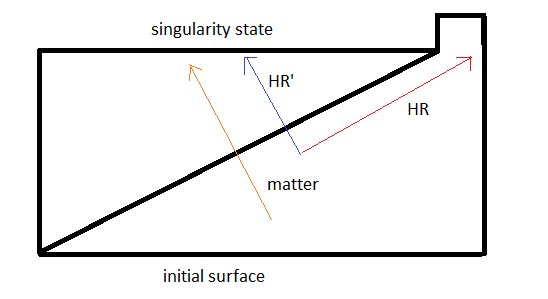}
\caption{semiclassical picture}
\end{figure}

	While energy conservation can often be preserved by careful choice of the decohering basis, the bleaching property introduces strong nonlocality by breaking entanglement with a local process. If one of a pair of entangled particles is sent into such a nonunitary process, it can be detected by local measurements on the other.  This will generically lead to loss of causality. A similar consequence follows from violation of the no cloning theorem as well as in many nonlinear generalizations of QM\cite{me,win}. 
The difference in observable consequences of information loss and hidden sector models can be quite strong. These models are most characterized by fundamental deformation of quantum mechanics where information is unrecoverable in principle.\\

	The "trace over" model is what most conservative proponents of information loss refer to as information loss. It is to simply take the trace over all observables on some boundary inside the horizon, possibly just surrounding the singularity. In this way each singularity in spacetime represents a sealed box of state information forever unavailable to regular observers. The accumulation of entropy comes from the gradual accumulation of entanglement with these 'un-observables'. As such, this scenario should belong in the hidden sector class of theories. One could argue that the thermofield double model is simply a convenient way to group all of these hidden pieces together.\\

\section{Hidden sector models}

	In unitary quantum mechanics any 'bleaching' process moves the entanglement to somewhere else. Hidden sector models put all in falling qubits into some inaccessible region. Stable remnants and baby universes would be examples of this type.\\

	Maldacena's thermofield double model falls into this type with the hidden sector here being highly nonlocal modes involving entanglement between left and right copies of the boundary field theories\cite{chow1}.  There are two troubling issues here. First, the thermofield double treatment can be used to model virtually any mundane lossy process, where information escapes into a thermal bath.  As such, the duality owes nothing in particular to the specific physics of black holes. Secondly, the image of left and right copies of the matter fields living on two sides of an Einstein-Rosen bridge does not seem to be physical. In a semi-classical model put onto the maximally extended metric, we simply have information being lost on both sides of the non-traversable wormhole. While matter may become entangled between one asymptotic region and another, the combined state is also still entangled with the singularity. This is sometimes combined with postselection so that all information ends up exchanged with the 'bath', leaving none on the singularity.\\

	All models of this type predict roughly the same physics, in which Hawking radiation is uncorrelated with infalling matter.  Entropy, at least as measured by local observers on our side of the black hole, increases monotonically for the entire evaporation process. While the overall state is pure, entanglement entropy is maximized with the 'second set' of fields.\\

	Combining postselection with the field doubling model gives another interesting theory, with it's own issues.  If one takes the full extended metric with second asymptotic region, and then imposes a final state on the singularity, we can time order the interior so that information that would have been lost is ‘reflected’ into the past of the second asymptotic region. This could also be seen as the second region acting as a unitary eraser on the infalling matter states. This scenario also has two problems. First it does not get the infalling states into the Hawking radiation. Both infalling matter and Hawking radiation are erased by their ‘mirror image’ from the second initial surface. This second initial surface acts as an arbitrarily large one time pad, like the thermal bath of thermofield dynamics, resulting in outgoing Hawking radiation that is uncorrelated with anything else from our visible asymptotic region. The dynamics observable from our asymptotic region would then be the same as any other generic hidden sector/loss model. Secondly, it requires that matter falling into the black hole from our side be already entangled with matter in this bath, even before the black hole forms. This may represent an unacceptable level of fine tuning of the initial state of the ‘second copy’ fields, while formally satisfying the unitary dynamics of ADS/CFT. \\

\begin{figure}
\includegraphics[scale=.3]{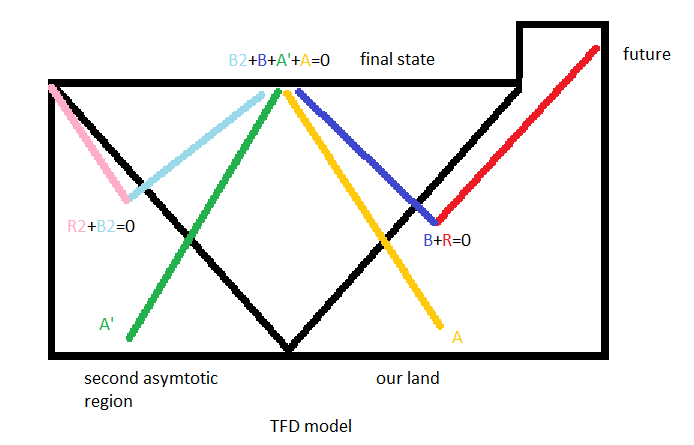}
\caption{thermofield double model}
\end{figure}

	A second paradox appears in the case of the thermofield model. Suppose the black hole is surrounded by a mirror. This reflects all the radiation from our region back inward, but has no effect on radiation emitted into the second asymptotic region. The hole in the second region will evaporate but the one in the first region will not. If the second region is a hidden sector, then both holes should vanish, taking the mass energy away into the hidden sector. On the other hand, if we started with a wormhole, the remaining mouth could be thought of as uncoupled or as a traversable time machine, depending on the completion of the interior. This is explored by Krasnikov\cite{kraz}.

\section{Super-selection/Post-selection}
   
	There have been a few proposals of this type\cite{mal1}. The black hole final state hypothesis is one, although a second could be constructed out of Hawking's cryptic comments about the Euclidean action suppressing black hole formation\cite{hawk06}.  In the first case we save information from hitting the singularity by using post-selection to place a final state boundary condition around the singularity. This boundary condition restricts the formation of Hawking radiation to only produce the negative energy modes that complement infalling matter so that the fields take on the correct values on a particular final time slice. Problems with this approach have been brought up\cite{pres2}, and it appears to lead to loss of causality.\\

	While the reverse causation aspect  has been used as a critique of these models, it has the advantage of making a possibly testable prediction. In a manner reminiscent of early big crunch scenarios, a generic post-selected field has a unique value on a final surface, and entropy increases on successive spatial slices at earlier rather than later times.  The generic case propagates backwards from the singularity, effectively choosing the advanced rather than retarded modes for infalling fields. The horizon acts as the turnover surface where retarded modes again begin dominating. If the turnover is sudden, a firewall may appear at the junction between the two regions. If it is gradual, or if smoothness of the junction is imposed, the transition would presumably be blurred over the duration of the collapse process, allowing some advanced modes to 'escape' to past infinity.  This escaped radiation might be observable using 'weak measurement' techniques, as a collapsing shell of red-shifted radiation that converges during black hole formation. The only advanced modes that would be excited far from the hole would be those eventually reaching the horizon. A normal dissipative measurement apparatus or absorber would appear to emit a contribution to the collapsing wave front rather than register a normal detection event. This is why weak or indirect measurement is needed. For a description of weak measurement and postselection physics, see\cite{ahr1}.\\

	The effect can also be understood to arise from the near saturation of the Bekenstien bound in the final moments of collapse. The time reversed movie of a system's evolution does not display much of an arrow of time if entropy is at or near maximum. A ten minute movie of mixing a drink may be indistinguishable from the reverse for nine out of the ten minutes.\\

	The other aspect of these models is the potential propagation of the postselected weighting factors onto the pre-collapsed matter.  The Hawing radiation acts as a thermal bath whose entropy is determined by the complement of the infalling state. In this way postselection may create a Boltzmann-like weighting factor for collapsing states, where the relative entropy between the collapsing star and final surface plays the role of a chemical potential. \\

	Just as post-selection deforms unitary quantum mechanics to create Hawking radiation correlations, a more extreme deformation may be used to prevent black holes from forming at all.\cite{hawk06} If a boundary condition is placed at the horizon we have a picture in which there is no spacetime interior at all, but a mirror-like firewall. The need for fields to know about global conditions in order to form such structures is satisfied by the nonlocal nature of post-selected systems. In each of these scenarios it is not clear how to treat more complex quantum states such as superpositions of black holes and regular spacetimes. The nonlinear nature of post-selection may lead to loss of unitarity anyway for these states.\\

\begin{figure}
\includegraphics[scale=.3]{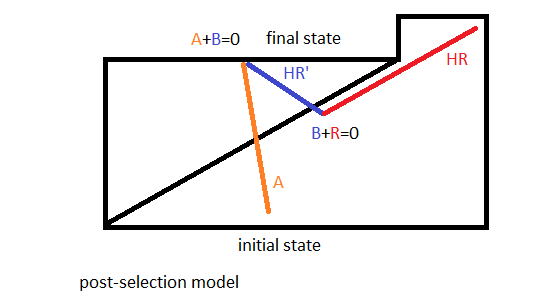}
\caption{black hole final state}
\end{figure}

\section{Information escape models}

\subsection{Tachyonic fields}
	Another approach is to explicitly add nonlocal or even acausal fields to the black hole interior to allow information to escape. Nonviolent nonlocality(NVNL) attempts to describe how a generic tachyonic field could leak the information while being strongly suppressed at large distances from the black hole\cite{gid1}. Whether this effective field arises from a more fundamental structure is another question. \\

	Effective source terms are chosen to comply with a no drama principle at least to first order, to give a sufficient channel for enough information to escape, and to suppress any acausal physics here. Source terms that have several powers of curvature in the coupling may be tuned to behave in this way, and often appear in modified gravity theories. This approach could also give a physical picture for information to escape into a tachyonic hidden sector. \\

	An alternative realization of NVNL would only allow communication between tachyonic and regular matter sectors where curvature was sufficiently large, or energy near the plank scale. Tachyons produced in high curvature areas would decay into particles of smaller negative mass and dissipate strongly in regular space, interacting only gravitationally, perhaps as dark energy. The tachyonic hidden sector could also be seen as a realization of a baby universe model since many locally positive energy modes in the baby universe may appear tachyonic when extended back into ours. 

\subsection{Non-black hole scenarios}
	Space-time in these scenarios is generally nonsingular. They may or may not have apparent horizons or long lived but temporary remnants. Some share the same problem with other horizon structure models that horizon formation is nonlocal. The main concern is then the deviation from classical gravity at low curvature, as well as perhaps significant changes to stellar physics. The messy business of finding string solutions that look like fuzzballs is a bit beyond the scope of this paper, but interesting questions can still be asked. Could stable stars form around fuzzballs or 'fire-bubbles'? Some models with nonsingular spacetimes involves long lived remnants and the back reaction of a negative energy flux into the hole\cite{hay}.  Singularity theorems are evaded by violating the energy conditions, with newer models replacing the long lived remnant with a bouncing black hole that explodes before the Page time. 

\subsection{ER=EPR}

	To understand this model, we can look at it as the natural extension of the hidden sector thermofield plus post-selection model. Consider a scenario with two asymptotic regions connected by a posts-selected singularity. All of the information of the infalling matter, as well as the entangled partners of Hawking pairs, can be ordered to end up in the second region. We don't want any losses from evaporation into the second region, and we do want to transfer the entanglement to outgoing HR.  ER=EPR attempts to fix these issues by identifying the ‘dumping ground’ surface with the states of outgoing HR, by using wormholes. Essentially it is cutting out the middle man. The second asymptotic region is replaced with a complex ‘hydra’ space-time that diverts the information to the mouths that are outgoing Hawking pairs. The final state boundary condition inside the hydra allows information to escape from the body(black hole) to the heads(HR). This is functionally equivalent to ordinary post-selection scenarios, with the added difficulty of the nontrivial interior 'hydra' space-time. It is hard to see what other predictions might be extracted from this proposal. The scenario of entanglement emergent spacetime should be seen as a distinct model from evaporation by post-selected wormhole pairs. The exploration of emergent spacetime models deserves a paper of it's own\cite{epr}.  
	
\begin{figure}
\includegraphics[scale=.3]{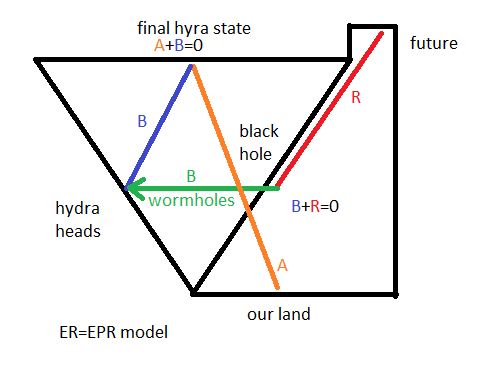}
\caption{hydra spacetime}
\end{figure}

\section{Complementarity}
	Complementarity is not a theory so much as a characteristic common to many models. 
These models attempt to have the best of both worlds, and are the main target of AMPS and similar arguments. Here we have observer dependent structure at the horizon. In one frame matter is deflected away by a stretched horizon. The AMPS argument and nice-slice argument are difficult problems since they consider frames that bridge the two complementary descriptions, attacking the basic premise that only individual observers' descriptions need to be consistent. If observers that could detect violations of unitarity can only do so in a transplankian regime, this is thought to be good enough. It's added to models with horizon structure to hide that structure from local observers. Fuzzball complementarity claims you do not notice hitting a fuzzball at low curvature, due to it having so many degrees of freedom. One could say the same thing about a baby universe theory, and call it 'baby complementarity', though this would be unnecessary. 

\section{Non-factorability}
	One comment made by Hawking shortly after his concession was that unitarity would be preserved by considering superpositions of spacetimes. Conversely he also said that information would still be lost when one is restricted to a particular classical black hole background. Most discussion of firewalls has taken place on such a classical background. The claim here is that the paradox can be circumvented by distributing the entanglement between early and late HR over a large number of geometries. By doing so it becomes exponentially harder for an observer outside the hole to reconstruct the copy of the emitted state needed to observe a violation of entanglement monogamy by crossing the horizon. Alternatively, we can have the Hawking pairs deviate from the normal statistics by a similar exponentially small amount, making the firewall that would result from such vacuum scrambling very weak. The small deviations and the eventual restoration of unitarity are left to quantum gravity. The primary assumption seems to be that all infalling states must be decohered with respect to a basis that spans the different background geometries, and that an observer who measures the entire state including such geometric correlations cannot fall through a 'particular' horizon\cite{stv}. Without quantum gravity to treat the superposition of classical geometries, it is difficult to analyze such a scenario. One observation here is that it predicts the apparent loss of correlation for the late time radiation by decohering it among large number of classical geometries.  

\section{Firewalls}
	This is most of the reason why the information paradox has seen a revival. An elegant thought experiment by AMPS, along the same lines as old no-cloning objections to complementarity led to the postulation of a new entity that would thwart any attempt to detect cloning inside the horizon. That would be called a firewall. Different models attempt to either avoid or generate firewalls in different ways. Blue shifted infalling matter from a second asymptotic region, does so in some models that have such regions. In string models such as Fuzzballs, it has been merged with the idea of the stretched horizon. In some others, it may be a consequence of the junction between interior and exterior fields. 

	Hawking radiation consists of vacuum pairs that should be maximally entangled upon creation. The outgoing part of that radiation should eventually be part of a pure state, so it appears that entanglement entropy of the black hole must at some point decrease to zero. This requires entanglement between radiation at different times or other matter entangled with the collapsing matter. According to AMPS, this violates the monogamy of maximal entanglement. To prevent this, everything is scrambled by the firewall before it can be observed.\\

	The principle of monogamy of entanglement results from the principle of no-cloning. Firewall and cloning paradoxes are essentially the same. Consider a black hole built out of one half of a large ensemble of EPR pairs.  Upon collapse entanglement entropy is already maximal, and should begin decreasing immediately with the emission of Hawking radiation. The mechanism whereby the horizon 'knows' when to form a firewall or emit quanta correlated with some newly arriving matter is still unknown. There seems to be no reason they should appear at any particular distance from the horizon, save that we don't want to be able to detect them. The formation of a radially static structure inside the horizon makes even less sense, as that would be a space like surface. The problem of apparent vs. global horizons makes it clear that they form from nonlocal effects. Lastly in models with an interior spacetime, an energetic firewall would appear to violate the field equations of GR.

\section{Summary}
A non-exhaustive table of popular models.

\begin{table}[h]
\begin{tabular}{lllllll}
     & Monotonic Entropy & Causality & Locality & Global Unitary & Firewalls** &  \\
TOS  & Y    &  Y      &  Y    &  N       & N &  \\
TFD  & Y    &   Y    &  Y      & Y*    & N &  \\
BHFS & N    & N      &   N    &  Y        & N &  \\
BU   & Y  & Y      &   Y    &   Y      & N &  \\
THS  &  Y    &    N    &  N     &   ?      & ? &  \\
NVNL & N    & N      &    N   &   ?      & ? &  \\
TFPS &    N  &     N   &   N    &   Y      & N &  \\
EPR  &  N  &   N    &   N    &  Y        & ? &  \\
FB   &    N  &    Y    &    N   &  Y       & Y &  \\
MSD  &    Y  &   Y     &  N     &  Y       & ? &  \\
REM &  N  & Y  &  Y  & Y  & N
\end{tabular}
\end{table}

TOS=trace over states ; TFD=thermofield double ; BHFS=black hole final state ; BU = baby universe ; THS = tachyonic hidden sector ; NVNL = nonviolent nonlocality ; TFPS =thermofield double plus postselection ; EPR = wormhole pair production model ; FB = Fuzzballs ; MSD = macroscopic entanglement decoherence; REM=remnants\\

	There are even more variations between these scenarios. One could imagine remnants that only partially restore purity but have shorter lifetimes. NVNL is a large class of potential models some of which might restore approximate locality and be unitary. * Unitarity is usually believed to follow from the ADS/CFT duality applied on a pre-collapse surface, combined with reversing the ordinary unitary collapse evolution. This is put to use explicitly in the pull back/push-forward method. ** There are versions with and without firewalls for TFD and EPR. One can also use a version of postselection to exclude firewalls specifically\cite{page}.\\

It seems that black hole physics is again in a golden age for theorists, but maybe just an age of headaches for everyone else.

\bibliographystyle{mdpi}
\makeatletter
\renewcommand\@biblabel[1]{#1. }
\makeatother

\end{document}